\documentclass[11pt]{article}
\usepackage{graphics, feynmp}

\newcommand{\be}{\begin{equation}}
\newcommand{\ee}{\end{equation}}
\newcommand{\bea}{\begin{eqnarray}}
\newcommand{\eea}{\end{eqnarray}}

\begin{document}
\begin{titlepage}
\def\thepage {}        

\title{Third Generation Seesaw Mixing with new Vector-Like Weak-Doublet Quarks}

\author{
Marko B. Popovic\thanks{e-mail address: 
markopop@buphy.bu.edu},\\
Department of Physics, Boston University, \\
590 Commonwealth Ave., Boston MA  02215}

\date{\today}

\maketitle

\bigskip
\begin{picture}(0,0)(0,0)
\put(295,250){BUHEP-01-1}
\put(295,235){hep-ph/0101123}
\end{picture}
\vspace{24pt}

\begin{abstract}

We present a class of models with a third generation seesaw mixing
with new vector-like weak-doublet quarks. We analyze a low-energy
phenomenology and present several strong dynamics, high-energy realizations of
these models. In the Topcolor type scenario, named Top-Bottom
Color, we obtain an effective, composite two Higgs doublet model where the
third generation isospin splitting is introduced via tilting interactions
related to the broken non-abelian gauge groups (i.e. without strong,
triviality-sensitive $U(1)$ groups). In addition, we discuss $t\overline{t}$
production at the Next Linear Collider (NLC) and suggest how one can
experimentally observe and distinguish between the weak-doublet and the weak-singlet types of seesaw mixing.

\pagestyle{empty}
\end{abstract}
\end{titlepage}



\section{Introduction and Synopsis}
\label{sec:intro}
\setcounter{equation}{0}

Particle physics of the last century gave us an impressively correct
low-energy phenomenological picture. A large share of that fundamental
development is the idea of the broken electro-weak (EW) symmetry explaining
the presence of heavy $Z$ and $W$ gauge bosons, the relation between their
masses, and their specific couplings to the fermions. However, the high-energy sector that overcomes the hierarchy problem and that is responsible for the EW symmetry breaking is yet to be understood.

Models with strong dynamics address the hierarchy problem through the slow,
logarithmic running of gauge couplings of non-abelian gauge groups that, after
a huge interval of energy scales, finally become strong enough to produce
substantial, ``interesting", effects on our low-energy world. Although
attractive, this idea poses many challenges - particularly in the
simultaneous dynamical EW symmetry breaking and mass generation for third generation quarks.

The general motivation for the week-singlet seesaw models
\cite{topseesaw1,topseesaw2} involving third generation quarks and
embedded in a Topcolor scheme \cite{topc} is to ``lower" the prediction for
the large dynamical mass of the top quark. If the compositness scale, the
scale of new physics, is not very large, then the dynamical mass is around
$600$ GeV - as implied by the Pagels-Stokar relationship \cite{Pages} (with the fixed
vacuum expectation value (VEV), $v_{EW}\approx 246$ GeV, of the composite
scalar field). With a seesaw mass mechanism with weak-singlet quarks, the physical top quark is just
partially dynamical and it can have the correct mass \cite{topseesaw1,topseesaw2}. In addition, the Topcolor scale is assumed to be low and the fine tuning problem in this context is resolved, although without a natural explanation of why the top Yukawa coupling equals one.

Earlier work in this direction addressed the seesaw mechanism with new
weak-singlet quarks in the top sector \cite{topseesaw1,topseesaw2}, as
well as in the bottom sector \cite{2see}. In the class of models presented in
this paper, we show that the realistic top and bottom quark masses may be
obtained via third generation seesaw mixing with new vector-like
weak-doublet\footnote{ It is often the case that a large number of extra
  $SU(2)_W$ doublets in models of strong dynamics may produce an unacceptably
  large contribution to the $S$ parameter. We show that such
  contributions are avoidable in the class of models that we consider here.}
quarks. With this setup it is possible to
economically\footnote{In this context ``economically" means a small number of
  additional ``new physics" parameters - mass parameters in particular -
  reflecting the high-energy dynamics.} produce the correct masses of the top
and bottom quarks in the Topcolor type of models and conserve the
low scale of Topcolor, with an acceptable amount of fine tuning. In
addition, in models that include simple, commuting (in respect to the weak
$SU(2)_W$ gauge group) Extended Technicolor (ETC) interactions \cite{etc}, we find
that the large mass of the top quark may be generated with slightly relaxed
corrections to the bottom quark EW coupling, which are otherwise proportional to the
large top quark mass \cite{Rb}. The low-energy phenomenological structure is
found to be just slightly altered from its successful Standard Model (SM) form. 

We are not first to introduce this extended third generation seesaw structure 
with weak-doublet quarks in the Topcolor type of models. An identical
low-energy structure with weak-doublet quarks \cite{topflavor} has been introduced in the spirit
of a third generation specific Topcolor-Assisted Technicolor TC$^2$
\cite{tc2} strong dynamical scheme (with identical extended Topcolor gauge
group sector but with top composite scalar field VEV $v_t\approx v_{EW}$). The main differences
between our strong dynamics, Top-Bottom Color, scenario and the one in
reference \cite{topflavor} are: $1)$ instead of using a strong $U(1)$ gauge
group - that necessarily has to be very strong to avoid a fine tuning problem
(and therefore, we believe, it is not easily accommodated in a natural
dynamical scheme) - we introduced the top-bottom mass splitting through the
effect of an additional strong, asymptotically free, non-abelian $SU(3)$
gauge group, and $2)$ we suggest the  bottom quark mass generation via the
same Topcolor type of mechanism while reference \cite{topflavor} suggests
that the bottom quark mass must be generated by different mechanism.

An additional motivation to study the third generation seesaw mechanisms with
weak-doublet quarks may be found in the recent development of models
with extra dimensions where the vector-like doubling of the fermion spectrum
arises naturally \cite{extra}.
    
First, we present the fermion charge assignments under the SM gauge group and
the mechanism of the mass mixing. Next, we analyze some of the
phenomenological consequences of this low-energy structure. Then, we explore
the possible high-energy realizations from which that low-energy structure
could emerge. Finally we discuss the direct search potential of $t\overline{t}$ production at the Next Linear Collider (NLC) and suggest how
one can experimentally observe and distinguish between the two (weak-singlet and weak-doublet) types of seesaw mechanisms.

\section{Low-Energy Weak-Doublets Structure}
\label{sec:LEW}
\setcounter{equation}{0}

We consider the following third generation fermionic representations under the SM gauge group, $SU(3)_{QCD} \otimes SU(2)_W \otimes U(1)_Y$
\bea
{\Psi_1}_L=\left( \begin{array}{c} t_L \\ b_L \end{array} \right)\, , \, (3,2,1/6); \; &
{\Psi_2}_L=\left( \begin{array}{c} T_L \\ B_L \end{array} \right)\, , \, (3,2,1/6); \nonumber \\ 
\Psi_R=\left( \begin{array}{c} T_R \\ B_R \end{array} \right)\, , \,
(3,2,1/6); \; & t_R \, , \, (3,1,2/3); \; \; \; b_R \, , \, (3,1,-1/3)\; .
\eea
As compared to the SM structure, the anomalies introduced by the additional left-handed and right-handed doublets cancel among themselves.

We assume the fermion mass mixing structure to be of the seesaw type. The possible dynamical origin of the specific mass terms that we introduce in this section and a motivation for their hierarchy will be discussed in Section 4 from the perspective of physics at higher energies.

First, we introduce the weak ``doublet - doublet" mass terms
\be
m_p{\overline{\Psi}_1}_L \Psi_R + m_q{\overline{\Psi}_2}_L \Psi_R \; .
\ee

Next, we introduce the ``doublet - singlet" mass terms
\be 
m_1{\overline{t}}_Lt_R + m_2{\overline{b}}_Lb_R \; .
\ee

Here we have assumed a hierarchy that requires $m_p, m_q > m_1, m_2$. Later,
we will explicitly consider the case where $m_p > m_q$ . We assume that no
other mass terms are present\footnote{The presence of
  $m_{1'}{\overline{T}}_Lt_R$ (or $m_{2'}{\overline{B}}_Lb_R$) mass term,
  with $m_{1'(2')} \ll m_1, m_2$, would only slightly alter the results of our analysis.}. 

The mass structure neatly splits into two separate structures for quarks of equal electric charge,
\bea
( \; \overline{t}_L \;\;\; \overline{T}_L \; ) \left( \begin{array}{cc} m_1 &
    m_p \\ 0 & m_q \end{array} \right) \left( \begin{array}{c} t_R \\ T_R
  \end{array} \right) , \\
( \; \overline{b}_L \;\;\; \overline{B}_L \; ) \left( \begin{array}{cc} m_2 &
    m_p \\ 0 & m_q \end{array} \right) \left( \begin{array}{c} b_R \\ B_R
  \end{array} \right) .
\eea
Performing the rotations separately on the left-handed and the right-handed states one diagonalizes the mass matrices and obtains the heavy (mostly non-SM)
and the light (mostly SM) mass eigenstates; for example
\be
T^m_R = - t_R \sin{\varphi_t} + T_R \cos{\varphi_t} \;\; , \;\;\; t^m_R = t_R
\cos{\varphi_t} + T_R \sin{\varphi_t} \; .
\ee 
Lower case fermions with the superscript $m$ denote the lighter mass eigenstates while upper case fermions with superscript $m$ represent the heavier mass eigenstates.

The physical (light and heavy) masses are given roughly as
\be
m_t \approx {{m_1 m_q} \over \sqrt{m_p^2 + m_q^2}} \;\; ; \;\; m_b \approx
{{m_2 m_q} \over \sqrt{m_p^2 + m_q^2}} \;\; ; \;\; M_T \approx M_B \approx M
= \sqrt{m_p^2 + m_q^2} \; .
\ee
The potentially dangerous rotations between right-handed weak singlet and doublet quarks (of the same electric charge) are roughly given by
\be
\sin^2{\varphi_t} \approx {{m_1^2 m_p^2} \over (m_p^2 + m_q^2)^2}
\;\; , \;\;\;\;\;  \sin^2{\varphi_b} \approx {{m_2^2 m_p^2} \over (m_p^2 +
  m_q^2)^2} \; ,
\ee
whereas the rotation among left-handed fermions (we include the superscript $L$ for distinction) are given by
\be
\sin^2\varphi^L_t \approx {m_p^2 \over {m_p^2 + m_q^2}} \;\; , \;\;\;
\sin^2\varphi^L_b \approx {m_p^2 \over {m_p^2 + m_q^2}} \; .
\ee
We adopt the convention that all $\sin\varphi > 0$, $\sin\varphi^L > 0$.

The couplings of the left-handed fermions to the $Z$ boson are the same as in the SM, while right-handed fermions' couplings to the $Z$ boson may be expressed in terms of the above angles as
   
\be
(\overline{t}^{\; m}_R \;\; \overline{T}^{\; m}_R)\left( \begin{array}{cc} T_3(t_L) \sin^2{\varphi_t}-Q(t)\sin^2{\theta_W} & -\cos{\varphi_t}\sin{\varphi_t}T_3(t_L) \\  -\cos{\varphi_t}\sin{\varphi_t}T_3(t_L) & T_3(t_L) \cos^2{\varphi_t}-Q(t)\sin^2{\theta_W} \end{array} \right) \left( \begin{array}{c} t^{\; m}_R \\ T^{\; m}_R \end{array} \right),
\ee
\be
(\overline{b}^{\; m}_R \;\; \overline{B}^{\; m}_R)\left( \begin{array}{cc} T_3(b_L) \sin^2{\varphi_b}-Q(b)\sin^2{\theta_W} & -\cos{\varphi_b}\sin{\varphi_b}T_3(b_L) \\  -\cos{\varphi_b}\sin{\varphi_b}T_3(b_L) & T_3(b_L) \cos^2{\varphi_b}-Q(b)\sin^2{\theta_W} \end{array} \right) \left( \begin{array}{c} b^{\; m}_R \\ B^{\; m}_R \end{array} \right),
\ee
with $ (e/\cos{\theta_W}\sin{\theta_W})$ factored out. As usual, $e$ symbolizes electric charge and $\theta_W$ is the weak angle. In order to have a viable model, we will show that $\sin^2{\varphi_b}$ should be very small.

The left-handed fermions' couplings to the $W$ boson may be expressed as

\[(\overline{b}^{\; m}_L \; \overline{B}^{\; m}_L)\left( \begin{array}{cc} \cos\varphi^L_t\cos\varphi^L_b+\sin\varphi^L_t\sin\varphi^L_b & \sin\varphi^L_t\cos\varphi^L_b-\cos\varphi^L_t\sin\varphi^L_b \\ \cos\varphi^L_t\sin\varphi^L_b-\sin\varphi^L_t\cos\varphi^L_b & 
\cos\varphi^L_t\cos\varphi^L_b+\sin\varphi^L_t\sin\varphi^L_b \end{array} \right) \left( \begin{array}{c} t^{\; m}_L \\ T^{\; m}_L \end{array} \right)\] 
\be
\approx (\overline{b}^{\; m}_L \; \overline{B}^{\; m}_L)\left(
  \begin{array}{cc} 1 & 0  \\ 0  & 1 \end{array} \right)
\left(\begin{array}{c} t^{\; m}_L \\ T^{\; m}_L \end{array}\right) \; , 
\ee
with $ (e /\sqrt{2}\sin{\theta_W})$ factored out. The right-handed fermions' couplings to the $W$ boson are
\be
(\overline{b}^{\; m}_R \;\; \overline{B}^{\; m}_R)\left( \begin{array}{cc} \sin\varphi_t\sin\varphi_b & -\cos\varphi_t\sin\varphi_b \\ -\sin\varphi_t\cos\varphi_b & 
\cos\varphi_t\cos\varphi_b \end{array} \right) \left( \begin{array}{c} t^{\;
  m}_R \\ T^{\; m}_R \end{array} \right) \; ,
\ee
again with $ (e/\sqrt{2}\sin{\theta_W})$ factored out. 

\section{Low-Energy Phenomenology}
\label{sec:LEP}
\setcounter{equation}{0}

The immediate low-energy physical implications of our low-energy setup with
small mixing angles in the right-handed fermionic sector are the
following: first, the light mass eigenstates corresponding to the top and bottom quarks have the correct $SU(3)_{QCD} \times U(1)_{EM}$ quantum
numbers; second, the EW couplings of the third generation quarks differ only slightly from those in the SM;
third, as we will show, the radiative corrections of interest resemble those
in the SM (and moreover we do not find $Z$-$\gamma$ mixing).
The oblique corrections place stronger constraint on the model parameters
than the tree-level limits coming from the $Z$-pole data (constraining only
$\varphi_b$). Left-handed mixings are not constrained by EW data. 

\subsection{Limits on $\varphi_b$ from the $Z$-pole data}
\label{sec:Limits}
 
From the $Z$ boson coupling to right-handed bottom quarks, equation (2.11), we note that the shift in the EW bottom coupling causes the parameter $R_b$ to decrease, i.e.
\be
{\delta R_b \over R_b^{SM}}=(R_b^{SM}-1){ 12
  \sin^2{\theta_W}\sin^2{\varphi_b} \over {9-12 \sin^2{\theta_W}+8
    \sin^4{\theta_W}}} \approx -33\% \sin^2{\varphi_b} \; ,
\ee
and as we see this change is proportional to $\sin^2{\varphi_b}$, clearly forcing the mixing angle to be small.
The one-loop corrections involving heavy $B^m (T^m)$ and $Z(W)$ exchange, are expected to be smaller due to the suppression via large $M_T, M_B$ masses ($\approx M$).

The parameter $R_b$ is not the only one influenced by the tree-level shift in the EW bottom coupling to the $Z$ boson.
We find that nine observables are affected: $\Gamma_Z$, $\Gamma_{had}$, $R_b$, $R_c$, $R_e$, $R_{\mu}$, $R_{\tau}$, $A_{FB}^b$, and $A_b$.  Using the general approach of reference \cite{fit}, and the experimentally measured and SM predicted values as in \cite{pdg}, we obtain
\begin{eqnarray}
& \; & {{\Delta \Gamma_Z} \over {( \Gamma_Z )}_{SM}}=-0.0628
\sin^2{\varphi_b} \; , \nonumber \\  & \; & {{\Delta \Gamma_{had}} \over {(
    \Gamma_{had} )}_{SM}} = - {{\Delta R_c} \over {(R_c)}_{SM}}= {{\Delta
    R_e} \over {(R_e)}_{SM}}= {{\Delta R_{\mu}} \over {(R_{\mu})}_{SM}}=
{{\Delta R_{\tau}} \over {(R_{\tau})}_{SM}}= -0.0899 \sin^2{\varphi_b} \; ,
\nonumber \\ & \; & {{\Delta R_b} \over {(R_b)}_{SM}}=-0.3268
\sin^2{\varphi_b} \; , \nonumber \\ & \; & {{\Delta A_{FB}^b} \over
  {(A_{FB}^b)}_{SM}} = {{\Delta A_b} \over {(A_b)}_{SM}}= 0.8630
\sin^2{\varphi_b} \; . 
\end{eqnarray}
We performed a fit to constrain the amount of mixing in the right-handed
bottom sector and found the $3\sigma$ constraint to be
\be
\sin^2{\varphi_b} \leq 0.0052 \; .
\ee
It is interesting to note that the fit generally prefers a negative values of $\sin^2{\varphi_b}$. For example, the SM value of $\sin^2{\varphi_b}=0$ is $2.1\sigma$ above the best fit value. 

The caveat is that in this calculation we included just the tree-level shift
in right-handed bottom $Z$ coupling, as given by equation (2.11), while we
neglected the effects of oblique corrections, implied by ``new physics", in
alteration of the $Z$-pole observable theoretical predictions. Clearly, once we
have the full model, including the entire scalar spectrum (and possibly the additional ``new physics") these corrections must be included.

Actually, in the framework of our low-energy setup, we expect $\sin^2{\varphi_b}$ to be much smaller than the limiting value in equation (3.3) . Just from the
consistency of the model, i.e. $\sin^2{\varphi_t} \leq 1$, we obtain
\be
\sin^2{\varphi_b} \leq {m_b^2 \over m_t^2} \approx 0.0006 \; ,
\ee
or an order of magnitude more stringent limit than in equation (3.3).

\subsection{Oblique Corrections}
\label{sec:Oblique}

The $(\delta T)_{NP} = {\alpha}^{-1}_{EM} (\delta \rho)_{NP}$ and $(\delta S)_{NP}$
oblique  radiative correction parameters represent the contribution of the 
``new physics" in our case, i.e. the contribution of the two-doublets structure minus 
the contribution of the SM one-doublet structure, to the differences that can be defined \cite{ST} as 
\bea
T={4 \over {{\alpha}_{EM} \; v^2_{EW}}} \left( \Pi_{11}(0) - \Pi_{33}(0) \right) \\
S=-{{16 \pi} \over M_Z^2}\left( \Pi_{3Y}(M_Z^2) - \Pi_{3Y}(0) \right) \; .
\eea
Here $\Pi_{jk}(q^2)$ are the gauge bosons' ($W_i^{\mu}$ and
$B^{\mu}$)  vacuum polarizations,  $M_Z$ is $Z$ boson mass, and $v_{EW} \approx 246$ GeV.

In the one-loop approximation, and in the basis of fermionic mass eigenstates, using the standard formalism (see, for example, \cite{ststandard}), from equations (2.7)-(2.13) we obtain   
\begin{eqnarray}
(\delta T)_{NP} & \approx & { {3 G_F} \over {8\pi^2 \sqrt{2} {\alpha}_{EM}}}
[ \; \sin^2{\varphi_t}\sin^2{\varphi_b} \; f(m_t, m_b) \nonumber \\
& + & \sin^2{\varphi_t} \; (\sin^2{\varphi_t}-\sin^2{\varphi_b})  \; f(m_t, M) \nonumber \\ 
& + & \sin^2{\varphi_b} \; (\sin^2{\varphi_b}-\sin^2{\varphi_t}) \; f(m_b, M) \; ]
\end{eqnarray}
\begin{eqnarray}
(\delta S)_{NP} & \approx & - { {2 {\alpha}_{EM} } \over
  {\sin{\theta_W}\cos{\theta_W} }} [ \; \sin^2{\varphi_t}\cos^2{\varphi_t} \;
\left( {5 \over 3} + \ln{m_t^2 \over M^2} \right) \nonumber \\ & + &
\sin^2{\varphi_b}\cos^2{\varphi_b} \; \left( {5 \over 3} + \ln{m_b^2 \over
    M^2} \right) \; ] \; ,
\end{eqnarray}
where $G_F$ is the Fermi constant and the function $f(x,y)$ is defined as 
\be
f(x,y) \; = \; x^2 + y^2 - {{2 x^2 y^2} \over {x^2 - y^2}} \; \ln{ x^2 \over
  y^2} \; .
\ee

In the limit $\sin^2{\varphi_t} , \sin^2{\varphi_b} \rightarrow 0$ we see that $(\delta T)_{NP} , (\delta S)_{NP} \rightarrow 0$ and therefore, with appropriately small mixing angles, corrections to the oblique parameters are easily kept under control.

In the previous subsection we found that $\varphi_b$ is expected to be extremely small and therefore we may safely neglect the terms with $\sin^2{\varphi_b}$.
This simplifies the expression for $(\delta T)_{NP}$, i.e.
\be
(\delta T)_{NP}  \approx { {3 G_F} \over {8\pi^2 \sqrt{2} {\alpha}_{EM}}}
\sin^4{\varphi_t} M^2 \approx {m_1^4 \over {m_p^2 m_t^2}} (\delta T)_t \; ,
\ee
where $(\delta T)_t$ is the SM contribution of the top quark and where we have assumed that $m_q$ is sufficiently smaller than $m_p$ so that $M \approx m_p$.

In the framework of the Topcolor scheme, if we take $m_1\approx 600$ GeV
(from the Pagels-Stokar relationship), then from  equation  (3.10) we find
that $m_p \approx M$ must be larger than\footnote{This result
  should be contrasted with the limit on the heavy fermions' masses,
  $\approx12$ TeV, in models with a seesaw mechanism with weak-singlet quarks
  in both, top and bottom, sectors \cite{2see}. Contrary to our case it is
  the parameter $R_b$ that place the most stringent limit in those models.} $\approx 5$ TeV. This sets $\approx 0.015$ as an upper bound for $\sin^2{\varphi_t}$. Therefore, from the model's consistency (equations (2.7)-(2.8)), one obtains $\sin^2{\varphi_b}$ to be smaller than $\approx 10^{-5}$. On the other hand, the ``new physics" contributions to the parameter $S$ are so small that they may be safely neglected.

It is interesting to note that in a slightly different variant of our model
in which the Yukawa mass terms, $m_1$ and $m_2$, link two right-handed singlets, $t_R$
and $b_R$, with both left-handed doublets, $\Psi_1=(t_L, \, b_L)^T$ and
$\Psi_2=(T_L, \, B_L)^T$, i.e.
\be
m_1{\overline{t}}_Lt_R + m_2{\overline{B}}_Lb_R \; ,
\ee
one obtains the mixing in the left-handed mass eigenstates sector satisfying
\be
\sin^2{{\varphi}_t^L} \approx \cos^2{{\varphi}_b^L} \; ,
\ee
instead of, as in our low-energy setup, $\sin^2{{\varphi}_t^L} \approx \sin^2{{\varphi}_b^L}$.
This in turn introduces off-diagonal terms in the $W$ coupling matrix and affects $(\delta T)_{NP}$ by an additional dangerous term, $ \approx (\delta T)_t (M^2/m_t^2) \sin^2({\varphi}_t^L - {\varphi_b^L})$. The cure for this, for example, would be to have $m_p \approx m_q$ which unfortunately makes the model structure highly constrained and therefore probably less viable.

\section{Weak-Doublets in Models of Strong Dynamics - High-Energy realizations}
\label{sec:toymoy}
\setcounter{equation}{0}

The high-energy physical implication of the low-energy setup in
Topcolor models, is that the weak-doublets structure offers a base
for an economical, simultaneous dynamical generation of the top and bottom
quark masses. In ``commuting" ETC models, the weak-doublet seesaw
mixing may relax the large shift in the bottom quark $Z$ coupling which is otherwise proportional to the large top quark mass. 

\subsection{Toy Modeling in the Topcolor Spirit - Top-Bottom Color}
\label{sec:toymoy1}

In this subsection, we sketch a strong dynamics scenario, named Top-Bottom Color, in which both the top and bottom masses are generated through Topcolor type of dynamics while they are directly linked to the bulk of EW symmetry breaking. This labeling reflects the fact that here instead of just one additional (top) color group we deal with the additional two (top-bottom) color groups. After introducing the gauge structure and the fermionic assignments under the additional gauge structure we concentrate on the dynamics of the Top-Bottom Color scenario.

As is traditional in the framework of topcolor models, one assumes that the
diagonal breaking $SU(3) \otimes SU(3) \rightarrow SU(3)$ provides the
attractive binding Nambu - Jona-Lasinio (NJL) interactions \cite{NJL} for
fermions transforming under a stronger ``initial" $SU(3)$ gauge group. If these NJL forces are appropriately strong, then the composite scalar field may acquire the non-zero VEV and furthermore, provide the dynamical masses for the strongly interacting fermions through their Yukawa couplings.

If the third generation quarks feel the same attractive NJL type of
interactions, then necessarily the condensation proceeds in an isospin symmetric
channel and consequently, the dynamical masses are equal. Therefore, in our class
of models, we expect that the top and bottom quarks must feel different NJL
forces. Since additional strong tilting $U(1)$  gauge groups are often found to be potentially dangerous \cite{schg} due to the triviality problem (see also for example \cite{pap1}), we turn our attention to additional $SU(N)$ gauge groups (with $N\geq3$) that exhibit more preferable high-energy behavior (characterized by a negative $ \beta $ function). Therefore, similar to \cite{topseesaw2,2see}, we consider a high-energy sector containing a ``trio" of strong $SU(3)$ copies, i.e. $SU(3)_1 \otimes SU(3)_2 \otimes SU(3)_3$, with strong gauge couplings, respectively, $g_1$, $g_2$, and $g_3$ as defined at some high-energy scale $\Lambda_1$. 

The strong multi-color sector exhibits a cascade of symmetry breakings of the form
\be
\begin{array}{c} SU(3)_1 \otimes SU(3)_2 \otimes SU(3)_3 \\ \nonumber

\downarrow \;\; [\Lambda_1, u_1] \;\; \downarrow \; \\ \nonumber
SU(3)_{1'} \otimes SU(3)_3 \\ \nonumber
\downarrow \;\; [\Lambda_2, u_2] \;\; \downarrow \; \\ \nonumber
SU(3)_{QCD} \end{array} ,
\ee
where $\Lambda_1$, $\Lambda_2$ are the scales of the appropriate symmetry
breakings and $u_1$, $u_2$ are the diagonal VEV's
of the effective EW singlet scalar fields, $\Phi_1$ and $\Phi_2$. In Appendix
A we discuss in detail the gauge bosons' mixing and couplings to the fermions
if the running of the strong gauge couplings is neglected (that algebraic structure may prove useful as a guideline for the features of the model that are more emphasized when we turn on the running of gauge couplings).

The low-energy setup fermions, introduced in Section 2, transform as
\bea
{\Psi_1}_L=\left( \begin{array}{c} t_L \\ b_L \end{array} \right)\, , \, (1,3,1,2,1/6); \;\;\;
{\Psi_2}_L=\left( \begin{array}{c} T_L \\ B_L \end{array} \right)\, , \, (1,1,3,2,1/6); \nonumber \\ 
\Psi_R=\left( \begin{array}{c} T_R \\ B_R \end{array} \right)\,  , \,  (1,1,3,2,1/6); \; \; t_R \, , \, (1,3,1,1,2/3); \; \; b_R \, , \, (3,1,1,1,-1/3);
\eea
where instead of gauge assignments under $SU(3)_{QCD}$, as in equation (2.1), we introduce assignments under the strong $SU(3)$ gauge ``trio". The schematic illustration of the cascade of symmetry breakings and  low-energy setup fermion charge assignments under the strong $SU(3)$ gauge groups is shown in Fig. 1.

\begin{figure}
\begin{center}
\rotatebox{0}{\scalebox{.5}{\includegraphics{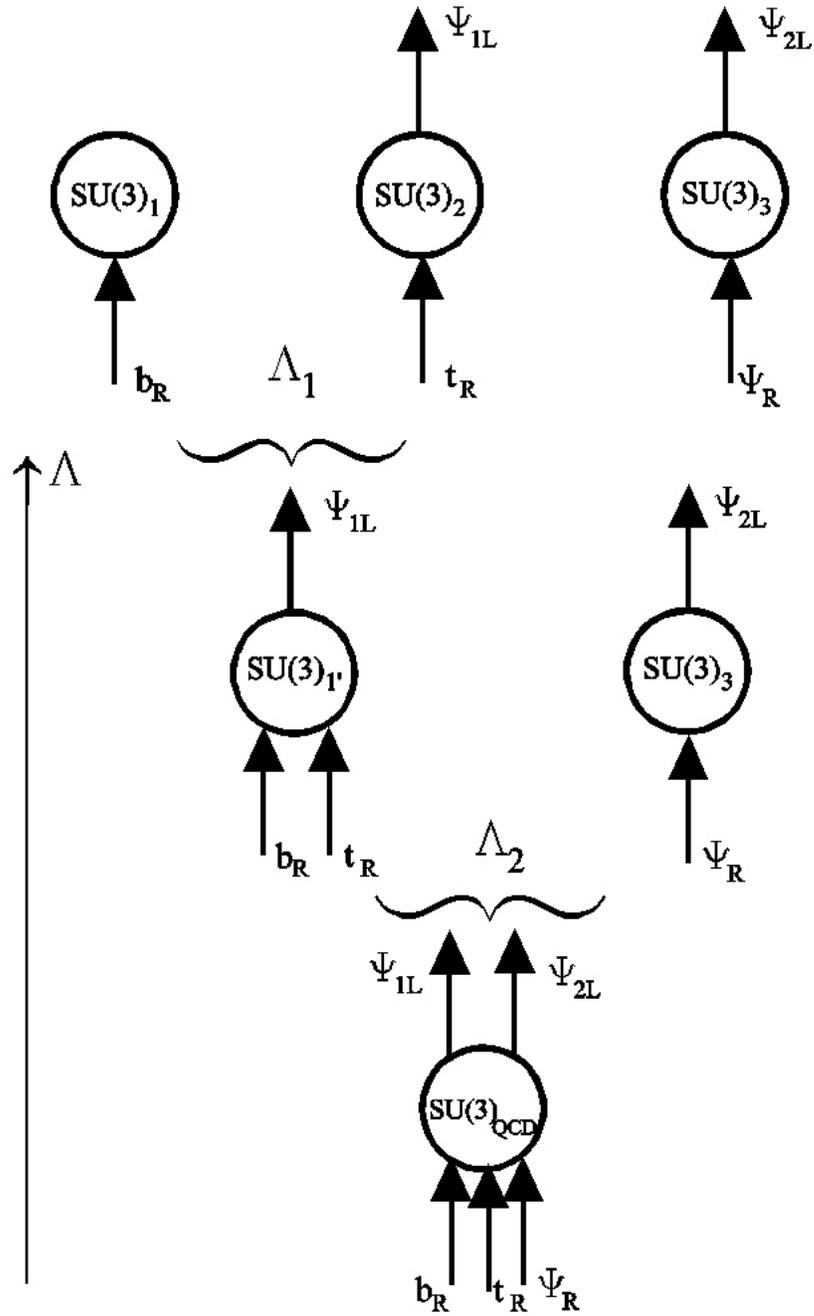}}}
\end{center}
\caption[lepton]{The schematic illustration of the cascade of symmetry
  breakings and low-energy setup fermion charge assignments under the strong
  $SU(3)$ gauge groups in the Top-Bottom Color scenario discussed in Section
  4.1. $\Lambda$ denotes an energy scale that increases in the direction of
  the long arrow.}
\label{Moose2}
\end{figure}

\begin{figure}
\begin{center}
\rotatebox{0}{\scalebox{.4}{\includegraphics{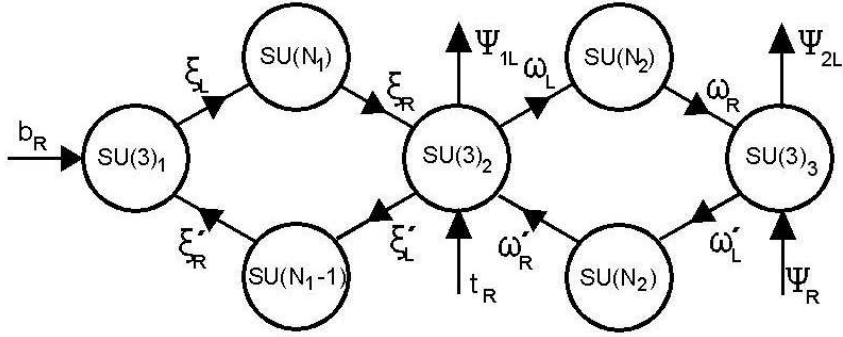}}}
\end{center}
\caption[lepton]{Illustration of the Top-Bottom Color model of Section 4.1 in ``Moose'' notation. Circles represent the gauge groups while lines represent fermions. Leaving lefthanded fermions (entering righthanded fermions) belong to the fundamental representation while entering lefthanded fermions (leaving righthanded fermions) belong to its conjugate representation under the appropriate gauge group.}
\label{Moose}
\end{figure}

With this assignment of the low-energy setup fermions, each of the $SU(3)_1$
and $SU(3)_2$ gauge groups suffers from gauge anomalies. In order to cancel
these anomalies and to provide the high-energy framework responsible for the
effective scalar fields, $\Phi_1$ and $\Phi_2$, and their diagonal VEV's, we
introduce four additional ``intermediate" gauge groups, $SU(N_0)$, $SU(N_1)$,
$SU(N_2)$ and $SU(N_3)$, plus two sets of ``intermediate" fermions,
weak-singlet $\xi$ and weak-doublet $\omega$ particles. This structure is
shown in compact ``Moose'' notation \cite{moose} in Fig. 2. 

The conditions for anomaly cancellations are
\be
N_0=N_1-1 \; , \;\;\;\; N_1 Y(\xi)=(N_1-1)Y(\xi')-{1 \over 3} \; ,
\ee
\be
N_2=N_3 \; , \;\;\;\; Y(\omega)=Y(\omega') .
\ee
In addition, we set $Y(\omega)={1 \over 6}$ and in order to have the preferred behavior of asymptotic freedom for ``intermediate" gauge groups we assume $N_1 \geq 4$ and $N_2 \geq 3$. All ``intermediate" fermions are degenerate fermions with the same isospin properties, and therefore, have no influence on the $S$ parameter.

The reason for the doubling of the ``intermediate" gauge groups in the $\xi$
sector is the cancellation of strong $SU(3)_1$ and $SU(3)_2$ anomalies
separately while avoiding the strong ``linking" $U(1)$ group. The reason for
the doubling of the ``intermediate" gauge groups in the $\omega$ sector is correct vacuum alignment giving the diagonal breaking of $SU(3)_2 \otimes SU(3)_3$ while preserving the cancellation of strong $SU(3)_2$ and $SU(3)_3$ anomalies separately. 
Apart from aesthetical reasons, we do not find the doubling of
``intermediate" gauge groups as a serious drawback for model dynamics. We
assume that while sliding down in energies the group $SU(N_1)$ gets strong
first at scale $\Lambda_1$ and triggers the diagonal breaking. An order of
magnitude or so below the scale $\Lambda_1$, the group $SU(N_1-1)$ becomes
strong and just minimally supplies the diagonal breaking VEV, $u_1$. In
addition, we assume that at approximately the same scale, $\Lambda_2$, both
$SU(N_2)$ groups get strong\footnote{This assumption about equal scales do
  not have to be unnatural - if both $SU(N_2)$ groups are embedded at some
  higher energy in a larger group, due to the identical running of the gauge
  coupling, as implied by the fermionic $\omega$ sector in Fig. 2, they become strong at approximately the same scale.}. 

The introduction of spectator (or light SM) quarks would probably simplify
this not so compelling gauge structure. Here however, for the purpose of
illustration, we concentrate just on the low-energy setup fermions. At this
point we are not concerned with the presence of various global $U(1)$ groups
that when broken would introduce strongly interacting Goldstone bosons - in a
more realistic model these may be taken care of via appropriate high-dimensional fermion operators that would give rise to the masses of the appropriate pseudo Nambu-Goldstone bosons in the multi-TeV range.     

At the energy scale $\Lambda_1$, the ``intermediate" group $SU(N_1)$ gets strong and binds the condensates of $\xi$ particles that break $SU(3)_1 \otimes
SU(3)_2$ down to its diagonal subgroup $SU(3)_{1'}$. This symmetry breaking generates an octet of massive precolorons, i.e. heavy colorons \cite{coloron}.
At energy scale $\Lambda_1$ we assume a hierarchy of the gauge coupling
strengths $g_2 > g_1 \gg g_3$ for the strong gauge ``trio". The $SU(3)_{1'}$ gauge coupling, $g_{1'}$, assumed large, is $g_{1'}= (g_1^{-2}+g_2^{-2})^{-{1 \over 2}}$. The precoloron exchange diagrams, below their mass, $M_c$, produce effective 4-fermion interactions.
Fiertzing interactions that couple lefthanded and righthanded currents of the
low-energy setup fermions in the large $N_c$ limit, one obtains NJL type interactions term proportional to
\be
{g_2^2 \over {M_c^2 (g_1^2 + g_2^2)}} \left[ g_2^2 \left(
    {\overline{\Psi}}_{1L} t_R \right) \left( {\overline{t}}_R \Psi_{1L}
  \right) - g_1^2 \left( {\overline{\Psi}}_{1L} b_R \right) \left(
    {\overline{b}}_R \Psi_{1L} \right) \right] \; .
\ee
However, we assume that these interactions are not strong enough to trigger dynamical condensation. Nonetheless, as we will see later, they represent a crucial tilting needed for isospin splitting of the third generation quarks.

At energy scale $\Lambda_2$ the ``intermediate" groups $SU(N_2)$ get strong and bind the condensates of $\omega$ particles that break $SU(3)_{1'} \otimes
SU(3)_3$ down to its diagonal subgroup $SU(3)_{QCD}$.  This symmetry breaking generates an octet of massive light colorons. The light coloron exchange
diagrams, below their mass, $M_L$, produce effective 4-fermion interactions.
Fiertzing interactions that couple lefthanded and righthanded currents of
low-energy setup fermions in the large $N_c$ limit, one obtains NJL type interaction terms proportional to
\be
{g_{1'}^4 \over { M_L^2 (g_{1'}^2 + g_{3}^2)}} \left[ \left(
    {\overline{\Psi}}_{1L} t_R \right) \left( {\overline{t}}_R \Psi_{1L}
  \right) + \left( {\overline{\Psi}}_{1L} b_R \right) \left( {\overline{b}}_R
    \Psi_{1L} \right) \right] \; .
\ee

Below the scale $\Lambda_1$, the strong coupling $g_{1'}$ runs enough so that
the interactions of equation (4.6), at the scale $M_L$, are above critical and trigger the NJL condensation. This interaction alone would
produce identical dynamical masses in the top and bottom quark
sectors. However, when combined with the tilting NJL interactions of equation (4.5) (that are somewhat amplified at the scale $M_L$) they provide the necessary isospin violation.
Therefore, at the same scale, just below the light coloron mass, dynamical condensation is triggered in both the top and the bottom sectors. Difference in the VEV's of top ($v_t$) and bottom ($v_b$) composite scalar fields is caused purely by the difference in the strength of the triggering NJL attractive interactions.

As the top and the bottom masses are linked through the same seesaw mixing, the prediction of the Top-Bottom Color scenario is
\be
{m_t \over m_b} \approx {m_1 \over m_2} \approx {v_t \over v_b} \; ,
\ee
and we obtain the effective composite two-doublet Higgs model with $\tan{\beta}=m_t/m_b$ at low-energies.

If the compositness scale, the scale of new physics, is not very large, then
the dynamical mass, $m_1$, is around $600$ GeV - as implied by the
Pagels-Stokar relationship (with fixed VEV, $v_t \approx v_{EW}\approx 246$
GeV, of the top composite scalar field). With a seesaw mass mechanism, the
physical top quark is just partially dynamical and it can have the correct
mass of $\approx175$ GeV.

One of the necessary ingredients of the seesaw mechanism, an $m_p \geq 5$ TeV mass term, may be thought to arise from an operator of the form
\be
{a \over M^2_{high}} ({\overline\Psi}_{1L} \omega_L ) ({\overline\omega}_R
\Psi_R) \; ,
\ee
where $a$ is a constant of order one and $M_{high}$ represents the high-energy
scale of the ``flavor" physics. Using the rules of naive dimensional analysis \cite{nda}, we find that this term provides the mass 
\be
m_p \approx 4 \pi a \left( u_2^2 \over M^2_{high} \right) u_2 \; . 
\ee
If flavor physics is weakly coupled,  one may have $M_{high}<M_L$; therefore without the need for introduction of the huge finely tuned gap between the masses $m_p$ and $M_L$ (as the scale of ``flavor" is naturally expected to be larger than $\Lambda_2$).
Although we would like to deal with the mass terms in the final instance, as if they were dynamical in their origin, to simplify the picture here, we may treat $m_q$ just as a bare mass term\footnote{That is, of course, not a necessity
  - one may consider, for example, additional ${{\omega}''}_L$ and
  ${{\omega}''}_R$ ``intermediate" fermions, that are weak doublets with
  hypercharge $1 \over 6$  and that link only the $SU(3)_3$ circle and, for
  example, the ``lower" $SU(N_2)$ circle in Fig. 2, i.e. transforming in the fundamental
  representation under these two gauge groups only. Below the scale ${\Lambda
    '}_2 (< \Lambda_2)$ (due to the slightly slower running the ``lower" $SU(N_2)$ become strong enough at slightly lower energy than its ``upper" $SU(N_2)$ partner group) their condensation may provide the $m_q (<m_p)$ mass
  term.}, where we assume $m_p > m_q$.

\subsection{Effect on $R_b$ in ``Commuting" ETC Models}
\label{sec:CoETC}
\setcounter{equation}{0}

In ETC models \cite{etc} in which $m_t$ is generated by the exchange of a weak-singlet ETC gauge boson, the shift in the EW bottom quark coupling is proportional \cite{Rb} to the large $m_t$. This alone causes the parameter
$R_b$ to decrease outside the experimentally allowed region. Armed with
additional left-handed and right-handed weak-doublets and our seesaw
mechanism, a similar analysis shows that this strong constraint may be
slightly relaxed.

As is standard \cite{Rb,standardRb}, one may consider the ETC gauge boson of mass $M_{ETC}$ coupling to the current
\be
-C{\overline{\Psi}_1}_L\gamma^{\mu}\chi_L+{1 \over
  C}{\overline{t}}_R\gamma^{\mu}U_R \; ,
\ee
where $\chi_L$ is the left-handed technifermion weak-doublet, $U_R$ is the
right-handed technifermion weak-singlet, and $C$ is a  Clebsch, expected to be of order one.

The mass $m_1$, related to the top quark mass, is generated through a four-fermion interaction (when one integrates out the heavy ETC boson at energies below $M_{ETC}$) of the form
\be
-{g^2_{ETC} \over M^2_{ETC}}\left({\overline{\Psi}_1}_L\gamma^{\mu}\chi_L\right)\left({\overline{U}}_R\gamma^{\mu}t_R\right) +h.c.
\ee
After Fiertzing this operator and using rules of naive dimensional analysis \cite{nda}, one estimates
\be
m_1 \approx {g^2_{ETC} \over M^2_{ETC}}4\pi v_t^3 \; ,
\ee
where $v_t = v_{EW}$.
On the other hand, the $b_L\overline{b}_L$ vertex to $Z$ is affected,
i.e. the tree-level $b_L$ coupling is shifted by \cite{Rb}
\be
\delta g_L^{ETC}(b_L)={1 \over 4} C^2 {m_1 \over 4 \pi v_{EW}}{e \over
  \sin{\theta_W}\cos{\theta_W}} \approx {1 \over 4} C^2
{{m_t/\cos{\varphi^L_t}} \over 4 \pi v_{EW}}{e \over
  \sin{\theta_W}\cos{\theta_W}} \; .
\ee
This large shift is translated to $b^m_L$ (mass eigenstate)
multiplied by $\cos^2{\varphi^L_b} \approx \cos^2{\varphi^L_t}$ and therefore the physical bottom coupling to $Z$ is affected by an amount proportional to $\cos{{\varphi}^L_t}$ (which is not constrained by the EW data, i.e. may be quite small)
\be
\delta g_L^{ETC}(b^m_L)= {1 \over 4} C^2 {m_t \over 4 \pi v_{EW}}{e \over
  \sin{\theta_W}\cos{\theta_W}} \cos{\varphi^L_t} \; .
\ee

Whether the described mechanism may open some new space for the simplest
class of models in which ETC and weak gauge groups commute is highly
questionable. Apart from the dangerous \cite{course} ETC ``direct"
\cite{deltaI2} and ``indirect" contributions to isospin violation in the $t-b$ sector (that exist independently of the introduced seesaw mechanism), raising the Yukawa type of mass term, i.e $m_1 \approx m_t / \cos{{\varphi}_b^L}$ necessarily 
lowers further the (already dangerously) low ETC top mass generation scale,
$M_{ETC} / g_{ETC}\approx 1$TeV \cite{course}, by a factor
$\sqrt{\cos{{\varphi}_b^L}} \approx \sqrt{\cos{{\varphi}_t^L}}$.   

\section{What can $t\overline{t}$ production at the NLC tell us?}
\label{sec:differ}
\setcounter{equation}{0}

Production of $t\overline{t}$ at the NLC (operating, for
example, at $\sqrt{s}=500$ GeV) is probably the most promising way to directly
observe top seesaw mixing and to differentiate between the weak-doublet and weak-singlet class of models. 

As we found in Section 3, the oblique $T$ parameter constrains the right-handed
top mixing, $\sin^2{\varphi_t}$, to be smaller than roughly $0.015$ in the
weak-doublet case. The right-handed top coupling to the $Z$ boson equals 
\be
{e \over {\cos{\theta_W}\sin{\theta_W}}}({1 \over 2}\sin^2{\varphi_t}- {2
  \over 3} \sin^2{\theta_W}) = {e \over {\cos{\theta_W}\sin{\theta_W}}}
(g_R^{SM}+\delta_D) \; ,
\ee
where $\delta_D=(1/2)\sin^2{\varphi_t}$, while the left-handed top coupling
is the same as in SM. This represents roughly a  $2\%$ maximal relative shift
in the top coupling. In the case of weak-singlet top seesaw mixing, the right-handed top coupling to $Z$ is unchanged from its SM form while the left-handed coupling equals
\be
{e \over {\cos{\theta_W}\sin{\theta_W}}} (-{1 \over 2}\sin^2{\varphi^s_t} +
{1 \over 2} - {2 \over 3} \sin^2{\theta_W}) = {e \over
  {\cos{\theta_W}\sin{\theta_W}}} (g_L^{SM} - \delta_S) \; ,
\ee
where $\varphi^s_t$ is the appropriate mixing angle in the weak-singlet case
and  $\delta_S=(1/2)\sin^2{\varphi^s_t}$. Similar to the weak-doublet case,
in the weak-singlet case \cite{topseesaw2, mypaper} $\sin^2{\varphi^s_t}$ is
found to be smaller than roughly $0.015$. This represents roughly a $4\%$ maximal
relative shift in the top coupling. Therefore, compared to the SM axial coupling, the axial couplings in both
doublet and singlet cases are smaller. However, the vector top coupling is
slightly larger in the doublet case and slightly smaller in the singlet case
as compared to the SM. 

Recently it has been suggested \cite{mixingextra} that the TESLA machine with
center of mass energy of $500$ GeV, an integrated luminosity of $300$
fb$^{-1}$, and unpolarized beams  may reach a precision of $2\%$ in the
determination of the $Z\overline{t}t$ coupling (from the total of almost
$35000$ top pair events in the channel with one $W$ decaying into $e\nu$ or $\mu\nu$ and
the other $W$ decaying hadronically). This may be enough for observing the
case with weak-singlet mixing but it is insufficient for the weak-doublet case.
We find the relative shifts in the total $t\overline{t}$ cross sections
relative to pure SM QED (in the above mentioned channel) to be
\be
{\Delta F_1 \over {(F_1)}_{SM}} = -0.06\,\delta_D \;\; \mbox{and} \;\;\; {\Delta F_1
  \over {(F_1)}_{SM}} = -0.44\,\delta_S \; ,
\ee 
for the weak-doublet and the weak-singlet cases respectively. 

Invoking the analysis of various asymmetries may considerably lower the
needed integrated luminosity and help in differentiating between the
weak-doublet and weak-singlet cases - for example by correlating the total cross section with forward-backward asymmetry. We find a relative forward-backward asymmetry (in the above mentioned channel) to be significant
\be
{{\Delta A_{FB}^t} \over {(A_{FB}^t)}_{SM}} = -1.60\,\delta_D \;\; \mbox{and}
\;\;\;
{{\Delta A_{FB}^t} \over {(A_{FB}^t)}_{SM}} = -1.46\,\delta_S \; ,
\ee    
for the weak-doublet and the weak-singlet cases respectively.
A more sophisticated analysis (with polarized beams in addition) allowing an even smaller integrated luminosity
may also be possible.

\section{Conclusions}
\label{sec:END}
\setcounter{equation}{0}

We presented a class of models with a third generation seesaw mixing
with new vector-like weak-doublet quarks. We analyzed a low-energy
phenomenology and proposed several strong dynamics, high-energy realizations of
these models.

We discussed a low-energy phenomenology and found that the oblique $T$ parameter places the strongest
constraints on the model parameters. Assuming the dynamical mass in the top
sector to be $600$ GeV, as determined by the Pagels-Stokar
relationship, we found that the heavy fermions must weight at least $5$ TeV. That implies an upper limit on the right-handed mixing in the top sector, $\sin^2{\varphi_t}$, to be
roughly $0.015$ (or $2\,\%$ relative shift in the  $Z\overline{t}t$
coupling) while the right-handed mixing in the bottom sector,
$\sin^2{\varphi_b}$, has to be smaller than $10^{-5}$. Left-handed mixings are not constrained by
 EW data. 

We proposed a class of high-energy models in the Topcolor spirit, named
Top-Bottom Color, with weak-doublets structure in the low-energy
limit. In this class of models we obtained an effective,
composite two Higgs doublet model with $\tan{\beta}=m_t/m_b$ at low-energy
and a realistic top quark mass. Instead of using a strong, triviality-sensitive $U(1)$ gauge group, we introduced the isospin mass splitting via tilting 4-fermion operators related to the broken non-abelian, asymptoticaly free gauge interactions.

In addition, we suggested that a search at the NLC and analysis of the
$t\overline{t}$ production asymmetries may have the best potential for direct observation
of different top couplings in the EW sector. Furthermore, we discussed
how one can distinguish between the two different (weak-doublet and
weak-singlet) types of seesaw mixing by correlating different measurements;
for example, the total cross section relative to the pure SM QED cross section
and the forward-backward asymmetry (although a more refined approach,
requiring even smaller integrated luminosity, may be possible as well).

\section*{ Acknowledgments }
\indent

The author would like to thank R.S. Chivukula, B.A. Dobrescu, H.-J. He, K.D. Lane,
K.R. Lynch, M. Postma, T. Rador and especially E.H. Simmons for useful
conversations and comments on the manuscript. This work was supported in part 
by the National Science Foundation under grant PHY-9501249, and by the
Department of Energy under grant DE-FG02-91ER40676.  

\appendix

\section{Details of Top-Bottom Color Symmetry Breaking (without running)}
\label{sec:app1}

To express some details of Top-Bottom Color symmetry breaking (without
running) in a compact form we use the following notation
\bea 
a=g_1 u_1 \; , \;\;\; b=g_2 u_1 \; , \;\;\; c=g_2 u_2 \; , \;\;\; d=g_3 u_2 \; , \\ \nonumber
N_0=\sqrt{a^2c^2 + a^2d^2 + b^2d^2} \; , \;\;\; N_1=a^2 + b^2 \; , \;\;\; N_2=c^2 + d^2 \; , \\ \nonumber
\mathrm{and} \;\;\; F(x)=N_2x^2+N_0^2(N_1-2x) > 0 \; , \;\; \forall x \in \Re
\; ,
\eea 
where gauge couplings ($g$'s) and VEV's ($u$'s) are as introduced in Section 4.1.

At the scale $\Lambda_1$ the groups $SU(3)_1$ and $SU(3)_2$ are broken down
to a diagonal $SU(3)_{1'}$. Therefore, the appropriate description, for energies below $\Lambda_1$ and above $\Lambda_2$, follows from the gauge boson mixing matrix
\be
\left( \begin{array} {cc} a^2 & -ab \\ -ab & b^2 \end{array} \right) \; .
\ee
Diagonalization of this matrix gives one zero mass eigenvalue, corresponding
to the pregluon field. The mass squared eigenvalue corresponding to the precoloron field is
\be
M_c^2=N_1=a^2+b^2=(g_1^2+g_2^2)u_1^2 \; .
\ee 
To obtain the precoloron and pregluon fields one rotates the initial unbroken
fields via the unitary real matrix
\be
\left( \begin{array} {cc} {a \over \sqrt{N_1}} & {-b \over \sqrt{N_1}} \\ {b
      \over \sqrt{N_1}} & {a \over \sqrt{N_1}} \end{array} \right) \; .
\ee
Therefore, the coupling of precolorons is read directly\footnote{As a
  consequence of the covariant derivative being unchanged.} as
\be
{a \over \sqrt{N_1}}g_1T_1 - {b \over \sqrt{N_1}}g_2T_2 = {{g_1^2 \,
    T_1-g_2^2 \, T_2} \over \sqrt{g_1^2 + g_2^2}} \; ,
\ee
where $T_1$ and $T_2$ are $SU(3)_1$ and $SU(3)_2$ generators. Similarly, the coupling of pregluons is
\be
{b \over \sqrt{N_1}}g_1T_1 + {a \over \sqrt{N_1}}g_2T_2 = g_{1'}(T_1 + T_2)
\; ,
\ee
where $g_{1'}= (g_1^{-2}+g_2^{-2})^{-{1 \over 2}}$.

For energies below $\Lambda_2$, the appropriate description follows from the gauge boson mixing matrix
\be
\left( \begin{array}{ccc} a^2 & -ab & 0 \\ -ab & b^2 + c^2 & -cd \\ 0 & -cd &
    d^2 \end{array} \right) \; .
\ee
Diagonalization of this matrix gives one zero mass eigenvalue, corresponding
to the standard gluon field. The squared masses of the heavy\footnote{Heavy
  coloron may be thought as low-energy descendant of the precoloron.} and the light colorons are given as
\be
{M^2_{H,\, L}}={{N_1+N_2} \over 2} \left( 1 \pm \sqrt{1-{4N_0^2 \over
      (N_1+N_2)^2}} \right) \; .
\ee
To obtain the heavy coloron, standard gluon, and light coloron fields one
rotates the initial unbroken gauge fields via the unitary real matrix
\be
\left( \begin{array} {ccc} {a\,b\,c \over \sqrt{F(M_H^2)}} & {(a^2-M_H^2)\,c
      \over \sqrt{F(M_H^2)}} & -{(N_1-M_H^2)\,d \over \sqrt{F(M_H^2)}} \\
    b\,d \over N_0 & a\,d \over N_0 & a\,c \over N_0 \\ {a\,b\,c \over
      \sqrt{F(M_L^2)}} & {(a^2-M_L^2)\,c \over \sqrt{F(M_L^2)}} &
    -{(N_1-M_L^2)\,d \over \sqrt{F(M_L^2)}} \end{array} \right) \; .
\ee
Therefore, the coupling of heavy colorons is 
\be
{a\,b\,c \over \sqrt{F(M_H^2)}}g_1T_1 + {(a^2-M_H^2)\,c \over
  \sqrt{F(M_H^2)}}g_2T_2 -{(N_1-M_H^2)\,d \over \sqrt{F(M_H^2)}}g_3T_3 \; .
\ee
The coupling of standard gluons is 
\be
{b\,d \over N_0} g_1 T_1 + {a\,d \over N_0} g_2 T_2 + {a\,c \over N_0} g_3
T_3 = g_{QCD}\,(T_1 + T_2 + T_3) \; ,
\ee
where $g_{QCD}=\left(g_1^{-2}+g_2^{-2}+g_3^{-2}\right)^{-{1 \over 2}}$. Finaly, the coupling of light colorons is
\be
{a\,b\,c \over \sqrt{F(M_L^2)}}g_1T_1 + {(a^2-M_L^2)\,c \over
  \sqrt{F(M_L^2)}}g_2T_2 -{(N_1-M_L^2)\,d \over \sqrt{F(M_L^2)}}g_3T_3 \; .
\ee 

If we assume a hierarchy with  $N_1 \gg N_2$ and particularly $ b^2 > a^2$, $c^2 > d^2$, and $a^2, b^2 \gg c^2, d^2$ roughly corresponding to the  scenario described in Section 4.1 with  $u_1^2 \gg u_2^2$ and $g_2^2 > g_1^2 \gg g_3^2$, a few simplifications are in order
\be
M_H^2 \approx N_1 + N_2 - {N_0^2 \over {N_1 + N_2}} \approx N_1 \;\;\;\;
M_L^2 \approx {N_0^2 \over {N_1 + N_2}} \; .
\ee
Therefore we find
\be
\sqrt{F(M_H^2)} \approx b c \sqrt{N_1} \approx b c \sqrt{a^2 + b^2} \; . 
\ee
Now, it is straightforward to see that the heavy coloron couples quite strongly just to fermions transforming under $SU(3)_2$, i.e.
\begin{eqnarray}
& \approx & {a \over \sqrt{a^2 + b^2}} g_1 T_1 - {b \over \sqrt{a^2 + b^2}} g_2 T_2 + {b c d \over (a^2 + b^2)^{3 \over 2} } g_3 T_3 \nonumber \\
& = & {{g_1^2 T_1 - g_2^2 T_2} \over \sqrt{g_1^2 + g_2^2}} + { g_2^2 g_3^2
  \over (g_1^2 + g_2^2)^{3 \over 2}} {u_2^2 \over u_1^2} T_3 \; .
\end{eqnarray}

On the other hand, in the same approximation, but for the small values of argument $x$, i.e. $x \ll N_1$, one has $F(x) \approx N_0^2 N_1$, and therefore, $\sqrt{F(M_L^2)} \approx a c \sqrt{a^2+b^2}$ .

This clearly implies that the light coloron couples strongly (with almost the
same strength) to the fermions transforming under both $SU(3)_1$ and
$SU(3)_2$ gauge groups. In other words, the light coloron coupling in this approximation may be expressed as
\begin{eqnarray}
& \approx & {b \over \sqrt{a^2 + b^2}} g_1 T_1 + {a \over \sqrt{a^2 + b^2}} g_2 T_2 -{ {d \sqrt{a^2 + b^2}} \over a c } g_3 T_3 \nonumber \\
& = & {g_1 g_2 \over \sqrt{g_1^2 + g_2^2}} (T_1 + T_2) - {{g_3^2 \sqrt{g_1^2 + g_2^2}} \over g_1 g_2 } T_3 \;\; \nonumber \\
& = & g_{1'} (T_1 + T_2) - {{g_3^2 \sqrt{g_1^2 + g_2^2}} \over g_1 g_2 } T_3 \;\; .
\end{eqnarray}

\end{document}